\documentclass{article}
\usepackage{PRIMEarxiv}
\usepackage[T1]{fontenc}
\usepackage[utf8]{inputenc}
\usepackage[numbers,sort&compress]{natbib}
\setcitestyle{round}
\usepackage{amsmath,amssymb,graphicx,longtable,array,calc,xcolor}
\usepackage{microtype}
\usepackage{placeins}
\usepackage{caption}
\usepackage[hypertexnames=false,hidelinks]{hyperref}
\DeclareUnicodeCharacter{00D7}{\ensuremath{\times}}
\DeclareUnicodeCharacter{2212}{\ensuremath{-}}
\DeclareUnicodeCharacter{2264}{\ensuremath{\leq}}
\DeclareUnicodeCharacter{03BD}{\ensuremath{\nu}}
\DeclareUnicodeCharacter{03C3}{\ensuremath{\sigma}}
\DeclareUnicodeCharacter{03B7}{\ensuremath{\eta}}
\DeclareUnicodeCharacter{0394}{\ensuremath{\Delta}}
\DeclareUnicodeCharacter{00B2}{\ensuremath{^2}}

\title{\normalfont Learning from fracture: Teaching language models to design through physical experience}
\author{
  \textbf{Bo Ni}$^{1,2,*}$\\[5pt]
  \normalfont $^{1}$Department of Materials Science and Engineering, Carnegie Mellon University,\\
  \normalfont 5000 Forbes Ave., Pittsburgh, PA 15213, USA\\
  \normalfont $^{2}$Generative Mechanics Laboratory (GenMech Lab), independent research initiative\\[5pt]
  \normalfont $^{*}$Correspondence: \href{mailto:nibo@cmu.edu}{nibo@cmu.edu}
}
\date{}
\hypersetup{pdftitle={Learning from fracture: Teaching language models to design through physical experience},pdfauthor={Bo Ni}}
\begin{document}
\maketitle
\begin{abstract}
How can language models learn to optimize physical designs through experience? We introduce FracLearn, an experience--abstraction--generalization framework evaluated through a deliberately unfamiliar, Tetris-inspired fracture game. The model first proposes architectures and examines simulated fracture responses to accumulate experience. Abstraction organizes these observations into conditional principles linking reinforcement placement, crack evolution and load transfer, with supporting evidence and failure conditions. Generalization challenges the model to combine these principles in new architectures and surpass the earlier design record. Campaigns demonstrate progressive refinement and recovery through hypothesis revision, yielding a best composite score 11.0\% above the experience-stage record. Fracture analyses connect improved performance to crack deflection and temporary trapping as toughening mechanisms, alongside greater residual load capacity. Learning occurs through external knowledge and decisions, without model-weight updates. This approach offers a route to scientific agents that turn computational experiments into reusable physical knowledge for materials discovery and engineering design.

\end{abstract}
\section{Introduction}

Nacre is made mostly of the brittle mineral aragonite. Yet its architecture, shaped by evolution, gives it substantial resistance to crack growth while carrying load \citep{jackson1988nacre,ji2004biological}. This ability to obtain damage tolerance from simple constituents has long inspired materials design \citep{chen2026nacre,gao2003flaws,wegst2015bioinspired,munch2008hybrid}. Advances in architected composites and additive manufacturing now extend that ambition, as demonstrated by damage-programmable metamaterials and toughened architected concrete \citep{gao2024damage,prihar2024concrete}. The expanding freedom to arrange matter allows engineers to pursue architectures far beyond familiar natural motifs in a broad design space \citep{schaedler2016architected,thompson2016additive}. It also brings a growing need for physical insight into arrangements for which established design rules offer little guidance and conventional trial-and-error strategies become less tractable or affordable.

Similar to nature, many engineering applications also call for materials that carry load with limited deformation, resist crack growth, and retain useful capacity after damage \citep{gupta2024cementitious}. These goals depend on many of the same architectural features, yet the mechanisms that support them can reinforce or compete with one another \citep{ritchie2011conflicts,li2024microstructured,evans2001nacre,barthelat2007mother}. Interactions among load transfer, interfacial failure, and inelastic deformation jointly influence how cracks advance \citep{barthelat2011toughness}; a material feature that promotes energy dissipation may also weaken a connection needed to sustain load \citep{barthelat2016interfaces,munch2008hybrid,he1989deflection,mirkhalaf2014glass}. Facing such coupling challenges, computational optimization and machine learning have explored these entangled interactions to discover improved architectures and reveal mechanisms behind their performance \citep{gao2024damage,gu2018hierarchical,li2024microstructured}. Learning from such designs offers a way to advance materials and fracture mechanics together: each tested architecture provides evidence about how resistance to failure emerges, and that understanding guides the next design.

For human researchers, learning to make such physical judgments involves years of education and research experience \citep{larkin1980expert}. Large language models (LLMs) open a distinctive possibility for expanding this capacity. An artificial system can formulate a design hypothesis and explain, in its own words, why it might work, making its proposed reasoning available for discussion, criticism, and revision \citep{wei2022chain,luu2024bioinspiredllm}. LLM agents have begun to formulate and solve mechanics and materials problems using simulation tools \citep{ni2024mechagents,ghafarollahi2025alloy,taherimousavi2026alloygen} and to plan and execute chemical experiments \citep{boiko2023autonomous,bran2024chemcrow}. Scientific-agent research further demonstrates the generation and refinement of hypotheses \citep{ghafarollahi2025sciagents}, with additional computation supporting improvements in this process \citep{gottweis2026coscientist}. Together with advances in allocating computation to LLM reasoning \citep{snell2025compute}, these developments raise the prospect of scaling physical reasoning itself: extending researchers' capacity to develop, compare, and revise physical ideas across a vast architectural design space, with explanations that can be interrogated and tested.

Can LLMs develop physical judgment through experience? Pretraining gives them access to extensive descriptions of the world, but knowledge from those records does not necessarily establish the consequences of a new intervention in a particular physical system \citep{bisk2020experience,bender2020nlu}. Evaluations in interactive physics simulations show that multimodal LLM agents can struggle to select actions that produce intended physical outcomes \citep{xu2026deepphy}. Studies also show that LLMs can rationalize incorrect answers and that unaided self-correction need not improve reasoning \citep{turpin2023unfaithful,huang2024selfcorrect}. For materials design, an explanation may be coherent in language while relying on assumptions that fail under load. Studies of human causal learning show how interventions and their consequences help distinguish competing explanations \citep{steyvers2003causal}. Research on human learning further connects reflection, comparison, and abstraction with the ability to use experience in new problems \citep{kolb1984experiential,chi1989selfexplanations,gick1983schema,gentner2003analogical}. Therefore, the open challenge is how to organize learning for LLMs so that physical outcomes inform the assumptions behind subsequent design decisions and the resulting lessons remain useful beyond the examples already encountered.

Research on embodied and experiential agents has begun to connect LLM decisions with environmental feedback \citep{yao2023react,ichter2023saycan,driess2023palme} and retained experience \citep{park2023generative}, including learning without weight updates \citep{shinn2023reflexion,wang2024voyager,zhao2024expel,vriza2026instruments}. Yet, scientific-optimization benchmarks show that an LLM agent can perform similarly with informative or shuffled feedback \citep{gupta2025bayesian}. These advances and limitations give new significance to how physical evaluations are used in design. In materials research, scaling the generation of hypotheses does not automatically scale the evidence needed to judge them. Experiments consume specimens and time \citep{lookman2019active}, while detailed fracture simulations face competing demands for fidelity and computational cost \citep{gu2018hierarchical,li2024microstructured}. Moreover, an improved response can reflect several interacting effects, leaving uncertain which lesson will remain useful in a new architecture. Learning is therefore constrained both by the experience available and by what can be inferred from it. The engineering opportunity is to connect the improvement of material architectures with learning to make physical judgments, so that each can support progress in the other.

\begin{figure}[htbp]
\centering
\includegraphics[width=0.75\linewidth,height=0.68\textheight,keepaspectratio]{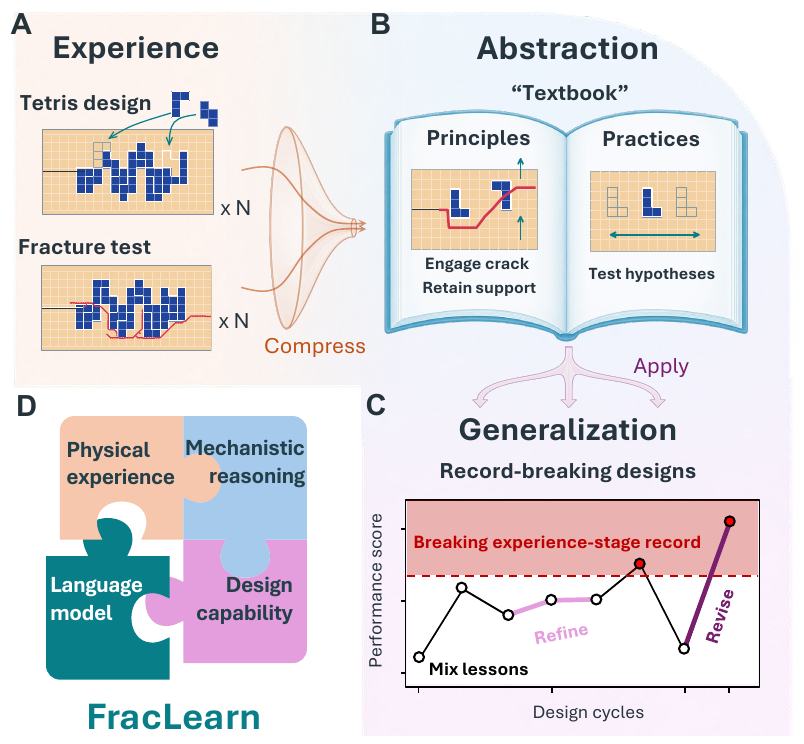}
\caption{\textbf{FracLearn: learning to design through physical experience.} The experience--abstraction--generalization framework is illustrated through a deliberately unfamiliar, Tetris-inspired fracture game. (A) Experience. Language model proposals are coupled with numerical search and fracture simulation to pursue focused optimization. Multiple campaigns, each centered on one design motif (×N), collectively generate diverse physical experience. Blue and peach denote reinforcement and matrix, respectively; red paths indicate cracks. (B) Abstraction. Experience across campaigns is distilled into a ``textbook'' of conditional principles and design practices, with supporting evidence and limits. (C) Generalization. The model is challenged to combine lessons in new architectures and surpass the experience-stage record. The schematic performance trajectory illustrates refinement and recovery through hypothesis revision. The dashed line marks the earlier record, and red points identify designs exceeding it. (D) FracLearn. The interlocking puzzle represents our learning strategy for aligning language model reasoning with physical experience, enabling mechanistic understanding and design capability to develop together. Demonstrated here through fracture design, this integration could extend to other physical systems, using the computational scalability of language-model reasoning to expand researchers' capacity to explore designs and develop explanations grounded in physical feedback.}
\label{fig:1}
\end{figure}

To address this coupled challenge, we develop FracLearn, a framework that organizes LLM learning through experience, abstraction, and generalization (Fig.~\ref{fig:1}). A deliberately unfamiliar, Tetris-inspired fracture game provides a model setting that couples LLM design proposals with numerical search and fracture simulation. We first show how campaigns focused on individual design motifs improve architectures while collectively building diverse fracture experience. We then examine how abstraction distills this experience into a textbook of conditional principles and practices, assessing their mechanical basis and limits. Generalization challenges the model to combine these lessons in new architectures, with refinement and hypothesis revision enabling performance beyond the experience-stage record. Alongside this learning progression, fracture analyses connect reinforcement placement to crack evolution and retained load-bearing capacity, relating design improvement to the mechanics underlying it. Learning proceeds through retained knowledge and evolving decisions, with model weights held fixed.

In the fracture prototype, FracLearn connects the search for improved architectures with learning how arrangements of simple constituents resist failure and retain load-bearing capacity. This joint pursuit of performance and understanding could support the development of damage-tolerant composites and additively manufactured structural materials, where architectural choices mediate competing demands (e.g., crack resistance and residual capacity). More fundamentally, FracLearn demonstrates a learning strategy for aligning LLM reasoning with physical experience, enabling mechanistic understanding and design capability to develop together. Extending this strategy to other physical systems could allow costly simulations and experiments to support knowledge that accumulates across design tasks, helping researchers investigate configurations beyond established design rules. Hypotheses and lessons expressed in language would remain available for examination and revision, with their value assessed through physical consequences. Coupling this accumulation of experience with the computational scalability of LLM reasoning offers a route toward expanding researchers' capacity both to discover better designs and to understand the physical principles behind them.

\FloatBarrier
\section{Results}

\subsection{A fracture-design game for learning from physical experience}

\begin{figure}[htbp]
\centering
\includegraphics[width=1.0\linewidth,height=0.68\textheight,keepaspectratio]{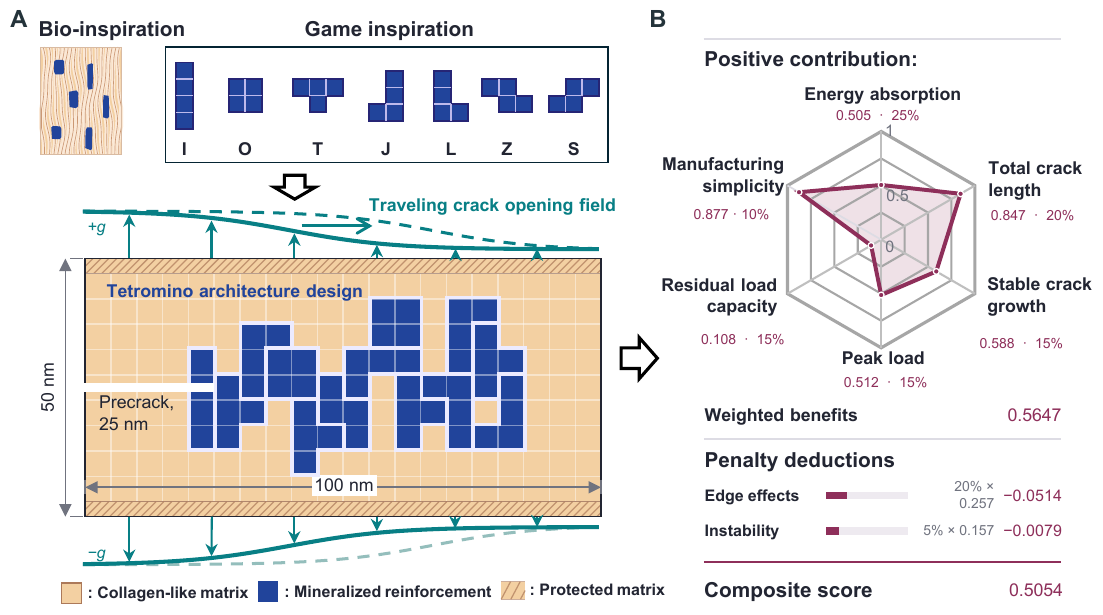}
\caption{\textbf{A fracture-design game connects architectural choices with physical feedback.} \textbf{(A)} A bioinspired material system with Tetris-based reinforcement assembly. Mineralized biological composites motivate the compliant matrix (apricot), stiffer and stronger reinforcement (ultramarine), and their elastic and fracture-property contrasts. Thirteen tetrominoes, drawn from seven shapes, occupy 52 cells of a 20 × 9 design grid. The architecture is tested in a 100 × 50 nm specimen with a 25-nm precrack. White outlines distinguish tetromino groupings; hatched matrix bands are protected against fracture. Opposing boundary displacements, +g and −g, impose a traveling opening field; solid and dashed profiles indicate successive positions. \textbf{(B)} Composite scoring combines six performance measures with two deductions. The polygon shows the normalized components, annotated with their values and weights. Weighted benefits of 0.5647, less deductions for edge effects and instability, give a composite score of 0.5054. Component definitions are given in Materials and Methods.}
\label{fig:2}
\end{figure}

We construct a two-phase material design game to investigate how LLMs learn to improve fracture resistance through physical experience (Fig.~\ref{fig:2}A). Inspired by mineralized biological composites \citep{ji2004biological}, the prototype combines a compliant matrix with stiffer reinforcement. Reinforcement-to-matrix ratios of 10 in Young's modulus and 6 in critical tensile stress for fracture make phase arrangement consequential for stress redistribution, crack propagation, and residual load-bearing capacity. By fixing constituent properties and loading, this model system isolates the influence of reinforcement architecture while retaining a broad range of possible crack--reinforcement interactions.

Reinforcement is assembled from tetrominoes, each comprising four connected cells (Fig.~\ref{fig:2}A). Thirteen pieces occupy 52 cells of a 20 × 9 design grid, fixing the material budget while allowing many combinations of shape, orientation, and placement. We deliberately introduce this construction rule as an additional challenge for learning. Although an LLM may have encountered composite design and Tetris separately during pretraining, their unconventional combination requires ideas about reinforcement and load transfer to be realized within tetromino-constrained architectures. Fracture simulation then tests whether the intended mechanical function survives this translation, probing the LLM's ability to adapt its design reasoning to the unfamiliar setting.

Each architecture is evaluated in a precracked specimen under a surfing boundary condition adapted from Hossain et al. \citep{hossain2014effective}. An opening profile translates along the upper and lower boundaries to promote sustained crack advance, while allowing the fracture path to respond to the phase arrangement (Fig.~\ref{fig:2}A). Finite-element simulations track crack evolution and load-bearing response. The fracture test was checked against single-phase benchmarks (fig.~\ref{fig:S1}; Materials and Methods).

Performance is assessed through six measures: energy absorption, total crack length, stable crack growth, peak load, residual load capacity, and manufacturing simplicity (Fig.~\ref{fig:2}B). Total crack length is estimated from deleted-element area, accounting for both the connected main crack and additional fracture. The mechanical measures capture potentially competing aspects of damage tolerance, motivated by the established difficulty of improving strength and fracture resistance together \citep{ritchie2011conflicts}. Manufacturing simplicity is represented by a geometric proxy based on phase interfaces and reinforcement connectivity. Normalized components are combined into a weighted score. Additional deductions are included for potential artifacts arising from the nonfracturing loading regions and for sudden large crack jumps, helping focus the search on physically meaningful architectural improvements. Retaining the individual score components exposes distinctions that the composite score can conceal: extensive cracking, for example, can coexist with little remaining load-bearing capacity. Together with crack paths and load histories, these measures supply both a design objective and physical evidence from which the LLM can learn.

\subsection{Learning through experience, abstraction, and generalization}

\begin{figure}[htbp]
\centering
\includegraphics[width=1.0\linewidth,height=0.68\textheight,keepaspectratio]{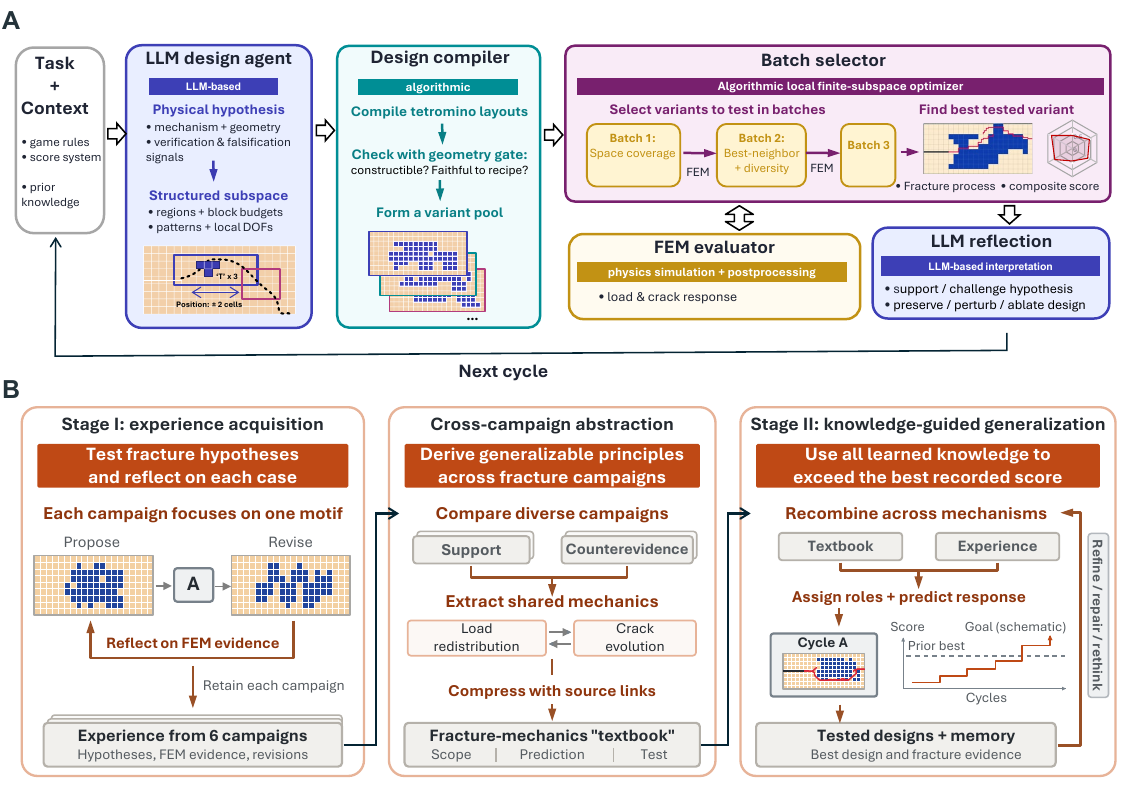}
\caption{\textbf{FracLearn organizes language-model learning through experience, abstraction, and generalization.} \textbf{(A)} A design cycle linking physical hypotheses to fracture evidence. The large language model (LLM) proposes a hypothesis, anticipated responses, and a structured design subspace. A deterministic compiler constructs admissible tetromino layouts, and a numerical selector explores this fixed pool through three batches of five finite-element method (FEM) evaluations. Crack and load responses and score components inform LLM reflection; the harness retains evidence and interpretations for the next cycle. \textbf{(B)} A learning progression connecting focused campaigns to new design tasks. Multiple cycles of (A) form a campaign. Six motif-focused campaigns generate diverse experience in Stage I. Cross-campaign abstraction compares supporting evidence and counterevidence, extracting conditional principles and practices into a textbook with source links, scope, predictions, and proposed tests. In Stage II, the LLM is tasked with combining textbook principles and retained experience in new architectures, assigning mechanical roles and predicting responses. Feedback-based instructions guide refinement, repair, or reconsideration toward exceeding the experience-stage record. Embedded A boxes refer to the shared design cycle. The score trajectory is schematic; the dashed line marks the earlier record. Blue and peach denote reinforcement and matrix; red paths indicate cracks. DOFs, degrees of freedom.}
\label{fig:3}
\end{figure}

To make fracture evaluations inform subsequent design reasoning, we build a computational harness around the LLM (Fig.~\ref{fig:3}A). It organizes a shared cycle of hypothesis formulation, numerical search, fracture evaluation, and reflection, using external feedback to address reported limitations of unaided LLM self-correction \citep{huang2024selfcorrect}. Given the game rules and available knowledge, the LLM proposes a physical hypothesis, anticipated fracture responses, and a design subspace defined by reinforcement regions, piece allocations, and allowed variations. A deterministic compiler realizes this subspace as admissible tetromino architectures and checks their consistency with the proposal.

Within each cycle, a numerical selector evaluates small groups of candidates, termed batches, using the fracture test and scoring system introduced above. The first batch samples the compiled family broadly; later batches combine neighbors of its best evaluated candidate with diverse alternatives. The family remains fixed during this search. The harness returns scores, component measures, and fracture diagnostics to the LLM, which compares the simulated responses with its expectations and identifies features to preserve or revise. Retaining these observations and reflections as context for the next proposal connects numerical exploration within a design family to revision of the hypothesis defining it.

Successive cycles linked by retained experience constitute a campaign. We organize these campaigns into an experience--abstraction--generalization progression, drawing on the connection between experience, reflection, abstract concepts, and further testing in human learning \citep{kolb1984experiential}. Each stage gives the physical evaluations a different learning purpose. During experience acquisition (Stage I), each campaign investigates one reinforcement motif (Fig.~\ref{fig:3}B, left). Six campaigns provide focused opportunities to improve architectures and confront their initial hypotheses with fracture responses, while collectively building diverse evidence for comparison.

Abstraction asks which mechanical lessons connect these distinct experiences (Fig.~\ref{fig:3}B, center). The harness assembles the campaign records and directs the LLM to compare supporting and conflicting observations, relate reinforcement arrangement to load redistribution and crack evolution, and condense these relationships into a fracture-mechanics textbook. Its conditional principles and design practices retain source cases, applicability limits, predicted responses, and proposed tests. This structure makes the basis of each lesson available for examination when it is carried into a new design.

Generalization (Stage II) tests the usefulness of these lessons beyond individual campaign motifs (Fig.~\ref{fig:3}B, right). Given the textbook and retained experience, the LLM is tasked with combining principles in new architectures, assigning their mechanical roles, and predicting the resulting response. The experience-stage record sets the performance target within the same material system, loading, and objective. Feedback-based instructions request refinement, repair, or reconsideration; the LLM selects the specific hypothesis and architectural changes. Further cycles therefore test both the performance of new designs and the physical expectations that guided their construction.

\subsection{Physical experience guides focused architectural refinement}

\begin{figure}[htbp]
\centering
\includegraphics[width=1.0\linewidth,height=0.68\textheight,keepaspectratio]{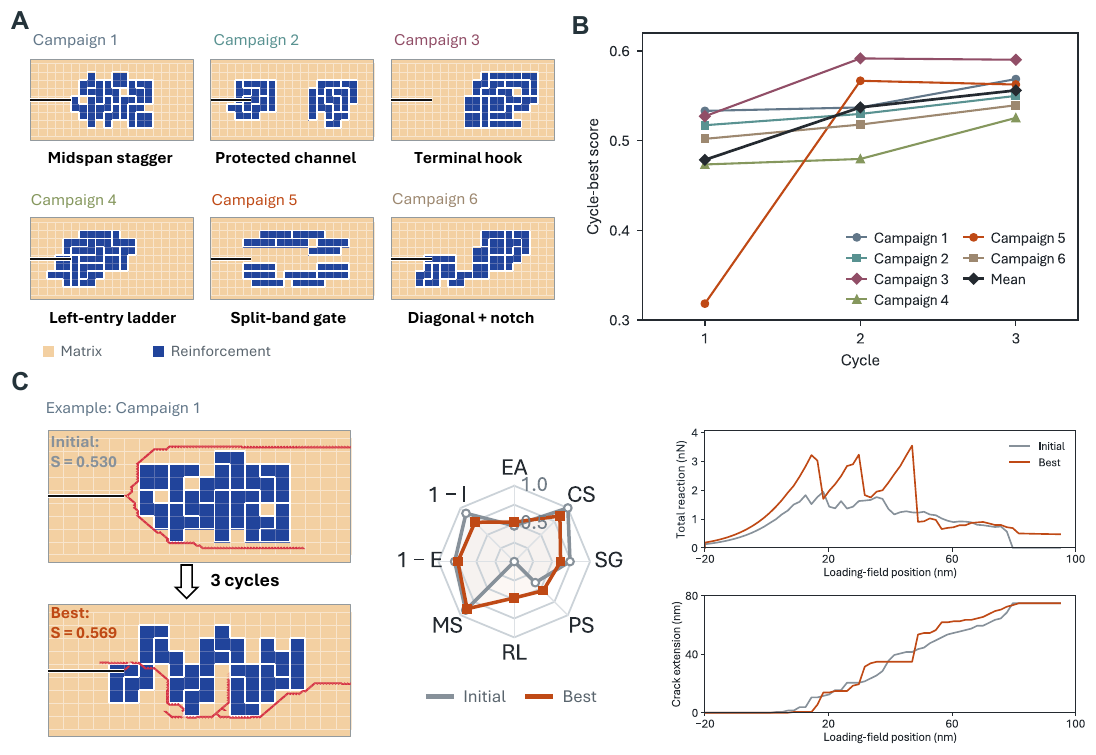}
\caption{\textbf{Physical experience guides motif-focused architectural refinement.} \textbf{(A)} Distinct reinforcement motifs explored in six campaigns. Each specimen is the geometric medoid of the 15 first-cycle designs. Peach and blue denote matrix and reinforcement; white outlines delimit tetrominoes. Black segments mark the 25 nm precrack in each 100 × 50 nm specimen. \textbf{(B)} Improvement through successive design cycles. Colored curves show cycle-best scores; the black curve shows their mean across campaigns, rising by 16.2\% from cycle 1 to cycle 3. Each cycle evaluates three batches of five candidates, totaling 270 evaluations. \textbf{(C)} Architectural refinement and its mechanical consequences in Campaign 1. The first-batch best (Initial, gray) is compared with the campaign best from cycle 3 (Best, orange). Architectures and final crack paths accompany component scores, total reaction, and connected-crack extension. Red traces mark deleted-element footprints. Polygon axes show energy absorption (EA), total crack length (CS), stable crack growth (SG), peak load (PS), residual load capacity (RL), manufacturing simplicity (MS), and the complements of the edge-effect (1 − E) and instability (1 − I) deductions. Values range from 0 to 1, with favorable values outward; weights and deductions are defined in Methods. Total reaction sums the magnitudes of the upper and lower boundary reactions. Connected-crack extension is measured from the initial crack tip. Histories are plotted against loading-field position.}
\label{fig:4}
\end{figure}

In the experience stage of FracLearn (Fig.~\ref{fig:3}B), six campaigns pursue focused refinement of distinct reinforcement motifs while collectively building fracture experience (Fig.~\ref{fig:4}A; table S1). Over three cycles per campaign, the system evaluates 270 architectures, and the mean cycle-best score rises from 0.4785 to 0.5560, an increase of 16.2\%. Every campaign finishes above its first-cycle best (Fig.~\ref{fig:4}B). The trajectories range from gradual gains to the pronounced improvement of the split-band motif in Campaign 5. Campaigns 3 and 5 nevertheless peak in the second cycle and decline slightly thereafter, showing that further revision does not guarantee improvement. The highest score, 0.5916, establishes the experience-stage record.

Campaign 1 illustrates how experience narrows an initially plausible design idea into a more selective strategy. Its initial hypothesis places staggered reinforcement near midspan to promote crack deflection and segmented advance. Comparisons within the first cycle show that stronger staggering or lateral shifts can improve energy-related measures while reducing connected crack advance or the composite score. The LLM consequently recommends, ``Do not globally increase stagger,'' and shifts toward local changes around successful arrangements. Later proposals preserve reinforcement across the crack corridor while adjusting its placement and extent. The retained reasoning thus connects observed trade-offs to specific choices about what to preserve and what to change.

Refinement also changes the fracture response (Fig.~\ref{fig:4}C). The campaign-best design, obtained in cycle 3, scores 0.569 compared with 0.530 for the first-batch best. In the initial specimen, cracks propagate around the upper and lower extents of the reinforced region, and the reaction eventually falls to nearly zero. The revised arrangement promotes crack deflection around successive reinforcement features. Intervals of limited crack advance accompany rising reaction, followed by load drops and abrupt extensions, consistent with temporary crack trapping followed by renewed propagation \citep{hossain2014effective}. Peak reaction increases from 1.94 to 3.55 nN (83\%), and the final reaction is 0.473 nN, or 13.3\% of the peak. Both specimens reach approximately 74.8 nm of connected crack extension; the revised architecture therefore retains load-bearing capacity after similarly extensive crack advance.

The component scores identify the mechanical trade-off behind this gain (Fig.~\ref{fig:4}C). Increased energy absorption, peak load, and residual capacity outweigh lower contributions from total crack length and stable crack growth. The architecture carries greater load but exhibits more intermittent crack advance. The LLM's retained reflections identify this loss of stability as a target for further local refinement. Experience therefore informs both a successful design and a mechanically specific next question.

Across the six motifs, focused refinement therefore generates both improved architectures and evidence linking reinforcement arrangement to fracture response. The next step asks which relationships recur across these experiences and can be expressed as conditional principles for design beyond an individual motif.

\subsection{Abstraction distills fracture experience into conditional design principles}

\begin{figure}[p]
\centering
\includegraphics[width=1.0\linewidth,height=0.68\textheight,keepaspectratio]{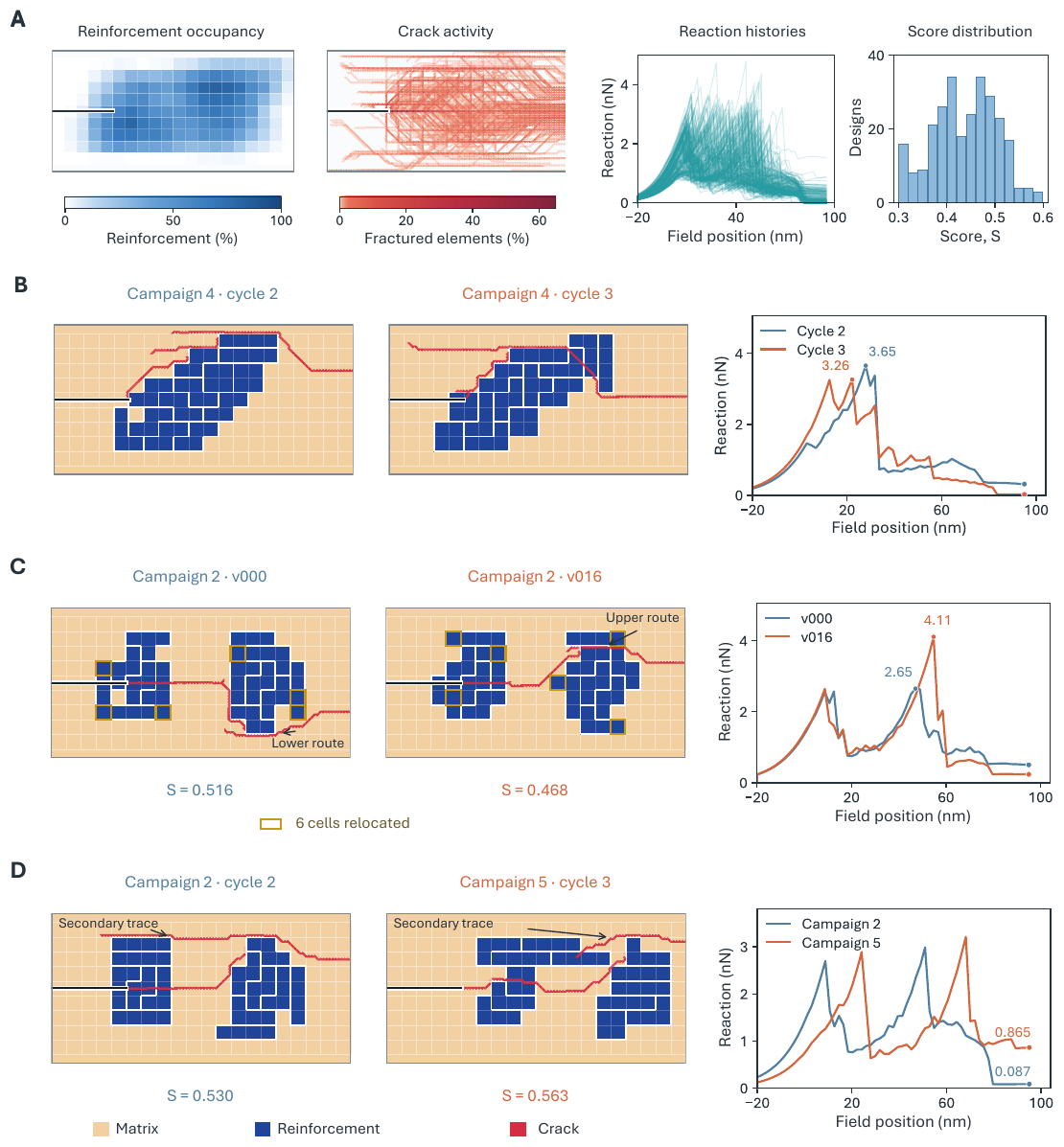}
\caption{\textbf{Abstraction connects diverse fracture experience to conditional design principles.} \textbf{(A)} Diverse physical experience. The 270 Stage I architectures are summarized by reinforcement occupancy, crack activity, reaction histories, and score distribution. Occupancy and crack activity give the percentages of designs containing reinforcement at each cell and a deleted element at each mesh location, respectively. \textbf{(B)} Crack--reinforcement interaction in Campaign 4 (movie S1). The cycle-2 specimen (left) shows main-crack arrest with separated fracture traces above the reinforcement band and a higher peak reaction; the cycle-3 specimen (right) develops a more continuous fracture route. \textbf{(C)} Sensitivity to reinforcement placement. Two Campaign 2, cycle-2 variants share 46 of 52 reinforcement cells. Gold outlines mark the six cells relocated between layouts. Fracture follows lower and upper routes around the downstream cluster, accompanied by different peak reactions. \textbf{(D)} Interpreting secondary fracture. Additional fracture traces accompany similar peak reactions but different final load retention. Blue and orange curves correspond to the left and right specimens. Peak reactions are labeled in B and C, and final reactions in D. Reaction is the sum of the magnitudes of the upper and lower boundary forces. S denotes the composite score. In specimen maps, blue denotes reinforcement, peach matrix, and red deleted elements; pale red marks deletions adjoining retained shared nodes. White lines delineate tetrominoes, and black segments indicate initial cracks.}
\label{fig:5}
\end{figure}

Abstraction organizes the diverse Stage I responses into lessons for designing and evaluating new architectures (Fig.~\ref{fig:3}B, center; Fig.~\ref{fig:5}A). Through cross-campaign comparison, the harness directs the LLM to consolidate experience into a textbook of four entries covering crack--reinforcement interaction, local placement, secondary fracture, and interpretation of mechanical evidence (table S2). We assess their mechanical basis below by connecting the recorded principles to representative fracture patterns and load responses.

The first entry emphasizes how reinforcement engages the advancing crack. In Campaign 4, an inclined reinforcement band deflects and arrests the main crack while separated crack segments develop above it (Fig.~\ref{fig:5}B, left). Peak reaction reaches 3.65 nN, compared with 3.26 nN in the right specimen, which develops a more continuous fracture route (movie S1). Additional element failure and a pronounced load drop occur while the tracked main crack remains arrested. These observations are consistent with crack arrest accompanied by daughter-crack formation, a mechanical interpretation informed by renucleation studies in heterogeneous materials \citep{hsueh2018renucleation}. The comparison shows how different crack--reinforcement interactions affect load resistance.

Realizing such interactions depends on local reinforcement placement, the focus of the second entry. Two designs share 46 of 52 reinforcement cells and the same two-cluster organization, yet relocating six cells switches fracture from the lower to the upper side of the downstream cluster (Fig.~\ref{fig:5}C). Peak reaction changes from 2.65 to 4.11 nN. This sensitivity supports the textbook's emphasis on actual spacing and alignment relative to the crack path: nominally similar architectures can engage reinforcement differently and redirect fracture, consistent with established crack-deflection and guidance mechanisms \citep{guo2020tstress,brodnik2020guiding}.

The third principle addresses the interpretation of secondary fracture. Two architectures develop additional fracture traces and reach similar peak reactions, yet retain final reactions of 0.087 and 0.865 nN (Fig.~\ref{fig:5}D). Secondary cracking thus accompanies markedly different residual responses. Additional fracture can dissipate energy while also interrupting load paths; its mechanical significance depends on where it develops and what load-bearing connections remain. This comparison supports the textbook\textquotesingle s practice of interpreting secondary cracks through their location and accompanying load response, giving mechanical meaning to features that cannot be judged by appearance alone.

The fourth entry organizes this assessment by checking admissibility, interpreting fracture paths, and considering the objective alongside its component measures. The preceding examples show why several observables are needed to connect a fracture pattern to its mechanical function. This reasoning is consistent with studies showing that neither greater crack deflection nor increased fracture-surface creation necessarily implies higher effective toughness \citep{abid2019fracture,hossain2014effective}.

The textbook therefore carries two complementary forms of knowledge into generalization: principles for arranging reinforcement and practices for judging their mechanical consequences. By linking proposed interactions to observable fracture and load responses, abstraction provides a basis for combining lessons across motifs and testing their usefulness in new architectures.

\subsection{Generalization advances fracture design beyond the experience record}

\begin{figure}[htbp]
\centering
\includegraphics[width=1.0\linewidth,height=0.68\textheight,keepaspectratio]{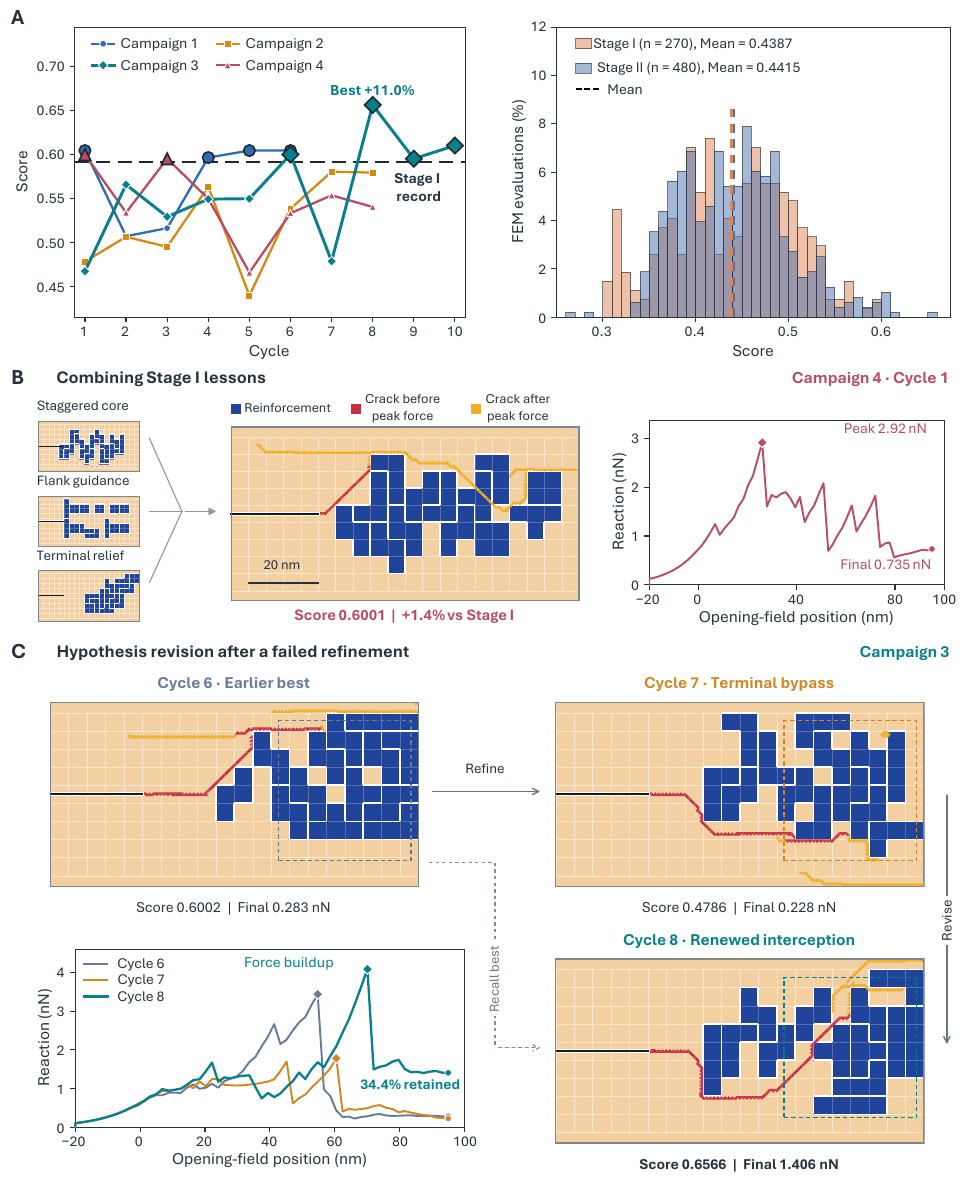}
\caption{\textbf{Generalization combines learned principles and revises hypotheses to advance fracture design.} \textbf{(A)} Performance beyond experience. Cycle-best scores for four Stage II campaigns (left); black outlines mark results above the dashed Stage I record. Distributions include all evaluations in each stage (right), normalized separately; dashed lines mark means. \textbf{(B)} Combining lessons in a new architecture. Stage I exemplars inform the staggered core, flank guidance, and terminal relief proposed for Campaign 4, cycle 1. The selected design exceeds the Stage I record and retains 25.2\% of peak reaction. Arrows indicate transfer of design ideas. \textbf{(C)} Recovery through hypothesis revision (movie S2). Campaign 3, cycles 6--8, compare an earlier best, an unsuccessful refinement with terminal bypass, and renewed crack interception after revision. The dashed arrow denotes recall of cycle 6; boxes locate terminal reinforcement. Cycle 8 exceeds cycle 6\textquotesingle s score by 9.4\% and retains 34.4\% of peak reaction. Blue and peach denote reinforcement and matrix. Red marks fracture accumulated through peak force, including the peak state; gold marks subsequent fracture, and black the precrack. Diamonds and circles identify peak and final reactions. Reaction sums the magnitudes of upper and lower boundary forces; the horizontal coordinate is the traveling opening-field position. Specimen maps use the undeformed geometry and preserve tetromino boundaries.}
\label{fig:6}
\end{figure}

In Stage II, the harness challenges the LLM to combine the abstracted lessons in architectures beyond individual Stage I motifs, within the same fracture-design problem (Fig.~\ref{fig:3}B, right). Three of four campaigns surpass the Stage I record (Fig.~\ref{fig:6}A). Mean candidate scores remain similar (0.4387 and 0.4415), whereas the best score rises from 0.5916 to 0.6566 (11.0\%). The campaign histories reveal three routes to this improvement: combining lessons at the outset, cumulative refinement, and recovery through hypothesis revision.

Combining lessons yields designs above the earlier record in the first 15-evaluation cycle of Campaigns 1 and 4. Campaign 4 illustrates how prior experience informs the initial proposal (Fig.~\ref{fig:6}B). Drawing on the textbook and retained examples, the LLM combines a staggered core, flank guidance from the split-band motif, and terminal relief informed by the hook motif to intercept and redirect the crack. Numerical search within this proposed subspace yields a best score of 0.6001 before any between-cycle revision. The crack deflects upward around downstream reinforcement, accompanied by repeated force drops and renewed loading. The specimen retains 25.2\% of its peak reaction after extensive fracture, coupling crack deflection with residual load-bearing capacity.

The recombined concepts also provide a basis for cumulative improvement. In Campaign 3, recombination followed by local refinement raises the cycle-best score from 0.5293 to 0.6002 over cycles 3--6, crossing the Stage I record (Fig.~\ref{fig:6}A). Generalization thus continues beyond an initial proposal: physical feedback guides the further development of an architecture assembled from earlier lessons.

Hypothesis revision converts a subsequent setback into progress beyond this refined design (Fig.~\ref{fig:6}C). In cycle 7, the crack turns downward and largely bypasses the terminal reinforcement, while peak reaction falls from 3.44 to 1.78 nN. The harness requests reconsideration and supplies both the failed trial and the successful cycle-6 design. The LLM attributes the setback to lost terminal interaction and proposes to ``Introduce a second compact overlap-support lobe.'' The revised subspace tests changes in terminal overlap and local reinforcement density. This response gives the placement principle examined in Fig.~\ref{fig:5}C a concrete design role: the next proposal seeks to recover crack--reinforcement engagement through a different arrangement.

The resulting cycle-8 architecture redirects the crack upward into the terminal reinforcement (Fig.~\ref{fig:6}C; movie S2). Reaction rises to 4.08 nN while tracked crack extension remains near 54.8 nm, before rapid advance and a force drop. This sequence is consistent with temporary crack trapping. After further fracture, the final reaction is 1.41 nN (34.4\% of peak), approximately five times that of cycle 6. The score improves by 9.4\% over that earlier best. Revision thus restores crack--reinforcement engagement while improving both the composite score and residual load-bearing capacity.

These cases show how knowledge distilled from experience can guide designs beyond its source motifs while remaining open to correction by subsequent physical outcomes.

\FloatBarrier
\section{Discussion}

FracLearn connects improvements in individual architectures to knowledge that can guide subsequent design. Focused campaigns generate evidence about particular design ideas. Abstraction makes this experience useful for new tasks by relating outcomes across motifs and retaining conditional principles and assessment practices. Generalization tests this guidance through new combinations. Campaigns 1 and 4 surpass the experience-stage record in their first cycle, showing how earlier lessons can provide a productive starting point for further learning (Fig.~\ref{fig:6}A, B). Learning here is expressed through retained knowledge and evolving design decisions, with model weights held fixed.

The content that transfers is a relationship between architecture and mechanical response. Relocating six of 52 reinforcement cells changes the fracture route and raises peak reaction from 2.65 to 4.11 nN within the same two-cluster motif (Fig.~\ref{fig:5}C). Additional fracture traces likewise accompany markedly different residual capacities (Fig.~\ref{fig:5}D). These comparisons give the textbook's principles their mechanical meaning: placement governs crack engagement, while the consequences of secondary fracture depend on the load-bearing connections that remain. Such relationships guide which interactions to preserve or reconsider in a new configuration.

Cyclic reflection tests how this knowledge is used in each new design. The harness uses performance and fracture diagnostics to request refinement or reconsideration, while the LLM chooses the next mechanical hypothesis and design subspace. In Campaign 3, this process led from terminal bypass to renewed crack interception and temporary trapping, increasing final reaction from 0.283 to 1.406 nN relative to the recalled design (Fig.~\ref{fig:6}C). The progressive stages broaden the knowledge available for design; the cycles test and revise its application. This emphasis on physical relationships and their recombination connects to broader proposals for causal and compositional learning in artificial intelligence \citep{lake2017machines}. Together, the stages and cycles connect accumulated experience to changes in the next physical experiment.

To assess the role of accumulated knowledge in this process, we performed a no learned-prior mechanics control. It retains the LLM--search--FEM workflow and within-campaign feedback but replaces learned mechanics guidance with generic task, scoring, and validity instructions (Materials and Methods). Its cycle-best scores decrease from 0.5265 to 0.5005 and 0.4911: the system produces admissible designs and revises them, yet does not improve its initial result. Over the same three-cycle window of 45 FEM evaluations, Stage II Campaigns 1--4 reach best scores of 0.6046, 0.5065, 0.5659, and 0.6001, respectively (Fig.~\ref{fig:6}A). Three exceed the control's best. Within this window, the comparison supports a role for supplied mechanics knowledge in design exploration and shows that feedback alone does not ensure productive revision.

The LLM makes the physical rationale directing the search explicit and revisable. Numerical search tests variants within a compiled subspace; the LLM can reconsider the mechanism and subspace pursued in the next cycle. Hybrid LLM--Bayesian optimization studies offer a related strategy for coupling language-based hypotheses with numerical search \citep{cisse2026closedloop}. This division of roles makes scientific claims available for researchers to examine, compare, and test. Generated explanations need not faithfully describe internal model reasoning \citep{turpin2023unfaithful,chen2025reasoning}, so their value here rests on whether their physical predictions withstand evaluation. FracLearn links those predictions to executable designs and simulated responses, allowing discrepancies to inform subsequent decisions.

The engineering opportunity is to increase what can be learned from each costly physical evaluation. In architected composites and additively manufactured metamaterials, a useful next test is whether the learned relationships can guide design when material contrast or loading changes, and whether the system recognizes when a lesson no longer applies. Such transfer would require fracture models or experiments that resolve the relevant failure processes; the present results concern a fixed two-dimensional computational setting. Scaling LLM reasoning could then expand the hypotheses considered, while accumulated physical evidence determines which lessons remain useful. This would make broader design exploration a means of building cumulative, testable mechanical knowledge.

\section{Conclusion}

FracLearn demonstrates how experience--abstraction--generalization can develop a language model's capacity to design through physical feedback. In the fracture prototype, combining earlier lessons, refining architectures, and revising unsuccessful hypotheses yield a best composite score 11.0\% above the experience-stage record. Fracture analyses connect these gains to reinforcement placement, crack deflection, temporary trapping, and retained load-bearing capacity. The learning process produces improved designs together with conditional mechanics principles and an explicit record of how physical evidence informs design decisions.

Such a capability could help convert the architectural freedom of additive manufacturing and mechanical metamaterials into control over damage and failure. Its engineering value would extend beyond a high-performing specimen to knowledge of which interactions to preserve as loading, material properties, or fabrication constraints change. That knowledge could support damage-tolerant materials designed around the coupled demands of crack resistance and load-bearing capacity.

More broadly, FracLearn offers a route toward scaling physical reasoning through knowledge that remains open to physical testing. We envision scientific agents that draw on accumulated experience to select informative experiments, distinguish competing mechanistic explanations, and develop design principles with researchers. In such a process, earlier physical outcomes would inform both the designs proposed and the questions asked next. Coupling scalable reasoning with this accumulation of evidence could expand the range of mechanisms accessible to scientific inquiry and the material functions accessible to engineering design.

\section{Materials and Methods}

\subsection{Material properties and design space}

The specimen was a 100 × 50 nm membrane of thickness 1 nm, with a 25 nm edge crack on its horizontal midline. A 20 × 9 interior grid defined the design space; protected matrix bands transmitted the boundary displacements. Thirteen nonoverlapping tetrominoes (I, O, T, J, L, S, and Z) occupied 52 reinforcement cells, with matrix elsewhere. Placement checks enforced piece geometry, the material budget, bounds, and protected regions. Adjoining reinforcement cells were mechanically continuous; tetromino outlines introduced no additional interfaces.

The constituent properties (Table 1) define a biologically inspired model composite \citep{ji2004biological}, without calibration to a specific tissue. Phase boundaries were perfectly bonded, with no separate debonding or plasticity law.

\textbf{Table 1. Constituent properties of the model composite.}

{\small\setlength{\tabcolsep}{4pt}\renewcommand{\arraystretch}{1.16}
\begin{longtable}{|
  >{\raggedright\arraybackslash}p{(\columnwidth - 6\tabcolsep - 4\arrayrulewidth) * \real{0.5231}}|
  >{\raggedright\arraybackslash}p{(\columnwidth - 6\tabcolsep - 4\arrayrulewidth) * \real{0.2385}}|
  >{\raggedright\arraybackslash}p{(\columnwidth - 6\tabcolsep - 4\arrayrulewidth) * \real{0.2385}}|}
\hline
\begin{minipage}[b]{\linewidth}\raggedright
\textbf{Constituent parameter}
\end{minipage} & \begin{minipage}[b]{\linewidth}\raggedright
\textbf{Matrix}
\end{minipage} & \begin{minipage}[b]{\linewidth}\raggedright
\textbf{Reinforcement}
\end{minipage} \\ \hline
\endhead
\endlastfoot
Young's modulus, E & 5 GPa & 50 GPa \\ \hline
Poisson's ratio, ν & 0.35 & 0.28 \\ \hline
Critical tensile stress for fracture, \emph{σ\textsubscript{c}} & 0.25 GPa & 1.5 GPa \\ \hline
\end{longtable}
}

Both phases used small-strain isotropic plane-stress elasticity, discretized with three-node constant-strain triangles on a center-cell mesh of nominal spacing 1 nm. Properties, mesh, and specimen geometry were fixed across designs.

\subsection{Fracture law and loading}

Element damage evolved irreversibly according to

\[d_{n} =  \max\left\lbrack d_{n - 1}, \max\left( 0, \frac{\sigma_{1,n}}{\sigma_{c}} \right) \right\rbrack,\]

where \(\sigma_{1,n}\) is the maximum principal stress entering the damage update and \(\sigma_{c}\) is the phase-specific critical tensile stress. Element stresses and strain energy were scaled by \(1 -  {clip\left( d_{n}, 0, 1 \right)}^{4}\), and elements were deleted when \(d_{n} \geq  1\). No independent Griffith surface-energy parameter was prescribed.

Adapting the surfing boundary condition of Hossain et al. \citep{hossain2014effective}, the vertical displacements were

\[u_{y}^{top} =  g(x, x_{c}),     u_{y}^{bottom} =  - g(x, x_{c}),\]

\[g(x, x_{c}) =  \frac{A}{1 +  \exp\lbrack(x -  x_{c})/d\rbrack}.\]

The one-sided amplitude A was ramped to 0.70 nm at \emph{x\textsubscript{c}} = −20 nm, with transition width d = 10 nm. At fixed amplitude, \emph{x\textsubscript{c}} advanced to 95 nm over 60 nominal quasi-static increments. Conjugate-gradient minimization followed each loading update, with adaptive subdivision where needed. The total opening was 2g; \emph{x\textsubscript{c}} denotes the imposed loading position, independently of the evolving crack tip.

\subsection{Response measures and composite score}

Reaction force was defined as \(F =  \left| F_{y}^{top} \right| +  \left| F_{y}^{bottom} \right|\), the sum of the magnitudes of the vertical reactions at the upper and lower loading boundaries. Peak and residual loads were the maximum and final accepted values of this quantity, respectively.

Connected crack extension was measured from the initial crack tip. Total crack length was estimated as \(L_{tot} =  A_{del}/b_{ref}\), where \(A_{del}\) is the deleted-element area and \(b_{ref} =  1.5 nm\) is the reference width used for scoring. The normalized component was \(CS =  clip\left( \frac{L_{tot}}{75 nm}, 0, 1 \right)\).

Energy consumed during fracture was

\[Q =  (W_{1} -  W_{0}) +  U_{0} -  U_{1},\]

where W is cumulative boundary work, U is stored element strain energy, and subscripts 0 and 1 denote the accepted ramp-end and final states. Boundary work was accumulated by trapezoidal integration of nodal reactions against imposed displacements. The energy input was \(G_{75} =  \max\left\lbrack 0, \frac{Q}{tL_{0}} \right\rbrack\), with thickness t = 1 nm and fixed initial ligament length \(L_{0} =  75 nm\), avoiding inflation from a short detected crack extension.

All designs used the same six normalized components and fixed reference scales. Let \(C(x) =  clip(x, 0, 1)\). For residual capacity, the diminishing-return map was

\[D_{k}(x) =  \frac{1 -  e^{- C(x)/k}}{1 -  e^{- 1/k}}.\]

Energy absorption (EA), peak load (PS), stable crack growth (SG), and residual load capacity (RL) were

\[EA =  \frac{G_{75}}{G_{75} +  G_{0}},     PS =  \frac{F_{\max}}{F_{\max} +  F_{P}},\]

\[SG =  \frac{1}{2}\left\lbrack \frac{N_{g}}{N_{g} +  24} +  1 -  P_{I} \right\rbrack,     RL =  \sqrt{D_{0.25}\left( \frac{F_{f}}{F_{\max}} \right)D_{0.35}\left( \frac{F_{f}}{F_{R}} \right)},\]

where \(N_{g}\) counts positive increments of connected-crack advance between accepted states, \(F_{\max}\) and \(F_{f}\) are peak and final reactions, and \(P_{I}\) is defined below. The reference scales were \(G_{0} =  0.01248301815 J/m²\), \(F_{P} =  3.0 nN\), and \(F_{R} =  1.872452722 nN\), expressed in the same physical units as the observables. Total crack length (CS) was normalized as defined above.

Manufacturing simplicity was

\[MS =  1 -  C\left( 0.60\eta +  0.40C\left( \frac{N_{c} -  1}{20} \right) \right),\]

where η is the fraction of horizontally or vertically neighboring design-grid cell pairs occupied by different phases, and \(N_{c}\) is the number of reinforcement clusters connected through shared cell edges. The two deductions used

\[P_{E} =  C\left( \max\left( \frac{f_{edge}}{0.12}, f_{sec,edge} \right) \right),     P_{I} =  C\left( \frac{{\Delta a}_{\max}}{70 nm} \right),\]

where \(f_{edge}\) is the near-boundary fraction among deleted designable elements, \(f_{sec,edge}\) is the corresponding fraction within secondary crack components, and \({\Delta a}_{\max}\) is the largest connected-crack increment. Both fractions used undeformed centroids ≤4.0 nm from any of the four exterior specimen edges, excluded protected nondesignable elements, and were zero for empty denominators. The weighted benefit and composite score were

\[B =  0.25 EA +  0.20 CS +  0.15 SG +  0.15 PS +  0.15 RL +  0.10 MS,\]

\[S =  clip(B -  0.20 P_{E} -  0.05 P_{I}, - 1, 1),\]

Nonfinite or nonpositive energy and force inputs were assigned zero contribution; zero peak force gave zero load retention. The instability measure contributes both to SG and to its direct deduction. Components, crack paths, and load histories were retained for interpretation; polygon area did not enter the score. The scoring implementation is provided in the \href{https://github.com/GenerativeMechanicsLab/FracLearn}{FracLearn repository} (Data and materials availability).

\subsection{Homogeneous fracture benchmarks}

Matrix-only and reinforcement-only specimens were tested with the same mesh and surfing protocol (fig.~\ref{fig:S1}). The energy consumed per projected crack area was

\[G_{bal} =  \frac{Q}{t \Delta a},\]

where t is specimen thickness and Δa is connected crack advance between the ramp-end and final states. A reference deletion-energy estimate was

\[G_{ref} =  \frac{{\sigma_{c}}^{2}}{2E}h_{eff},     h_{eff} =  \frac{a}{\sqrt{2}},\]

where a = 1 nm is the mesh spacing and the same effective deletion-band width was used for both phases. Both specimens developed essentially straight through-cracks with 74.83 nm connected extension. The measured-to-reference energy ratios were 1.049 for the matrix and 0.943 for reinforcement. The reinforcement-to-matrix energy ratio was 3.24, compared with the reference prediction of 3.60. These benchmarks assess consistency at the selected mesh and loading protocol; they do not establish mesh-independent material toughness.

\subsection{Harness-mediated feedback and learning tasks}

The design agent used the OpenAI Responses API with model alias gpt-5.4 and service-default settings. Proposals and reflections were stored as structured records, including hypotheses, anticipated FEM signatures, and falsification criteria. One schema-repair call was permitted for an invalid proposal. Model weights remained fixed. Source code, prompts, campaign records, and replay notebooks are provided in the FracLearn repository (Data and materials availability).

The LLM specified one to three reinforcement regions, regional tetromino budgets and patterns, and at most two design degrees of freedom. A deterministic compiler with base seed 17 instantiated exact layouts and checked geometry and recipe consistency. The numerical selector sampled the fixed pool in batches of five. Initial selection maximized coverage in the allowed degrees of freedom and cell occupancy; subsequent batches selected at most two unevaluated neighbors of the best tested design and filled the remainder by diversity. All reported cycles completed three batches, totaling 15 FEM evaluations; LLM revisions of the subspace occurred between cycles.

Before each call, the harness assembled task instructions, available knowledge, and retained campaign context. Scores, component measures, fracture diagnostics, and LLM reflections were compressed into external memory for later decisions \citep{shinn2023reflexion,zhao2024expel}. Human audit prose and the researcher-facing high-score archive were excluded from agent context. The retained record therefore documents the LLM's stated hypotheses and interpretations.

During generalization, predefined rules used performance changes and fracture-diagnostic flags to request refinement, repair, or reconsideration. These instructions specified the required extent of revision; the LLM chose the mechanical hypothesis and corresponding architectural changes. Checks assessed whether proposals met the requested structure and geometric constraints. Such checks and rule-based summaries governed the workflow; mechanical interpretations were assessed separately against the simulated fracture response.

\subsection{Experience acquisition campaigns}

Six Stage I campaigns used the shared design cycle. Initial context included 12 FEM observations from the same fracture setup, using unrestricted random clusters with 54 reinforcement cells. These were supplied as qualitative physical evidence, not tetromino templates or named principles, and were separate from the 270 reported Stage I evaluations using thirteen tetrominoes and 52 reinforcement cells. Between campaigns, accumulated mechanism memory could be retained while a diversity directive requested a distinct motif. Thus, the campaigns were not memory-isolated replicates. Initial hypotheses are summarized in table S1.

Each campaign comprised three cycles, with 20 admissible variants compiled and 15 evaluated per cycle. The LLM revised the design family between cycles while retaining the campaign motif. Instructions required previously successful features to be preserved or their alteration to be justified as a test; geometry checks requested repair when necessary. Campaign lengths and continuations were specified by the research protocol.

\subsection{Performance and fracture comparisons}

A cycle-best score was the maximum among that cycle's 15 evaluations; the mean trajectory in Fig.~\ref{fig:4}B was the arithmetic mean across the six campaigns. The reported 16.2\% gain was calculated from the first- and third-cycle means. The experience-stage record was the maximum over all 270 evaluations. These descriptive comparisons characterize the coupled LLM, search, and simulation workflow; the six campaigns explore different motifs and are not replicate runs of a single hypothesis.

Figure 4C compares Campaign 1's first-batch best with its campaign best. Its initial score therefore differs from the first-cycle best in Fig.~\ref{fig:4}B, which includes two further batches. Crack and reaction histories use the translating loading-field position as their common coordinate. The quoted retained fraction is the final-to-peak reaction ratio. Component polygons use the same score definitions as Fig.~\ref{fig:2}; edge-effect and instability deductions are displayed as 1 − E and 1 − I. Recorded LLM interpretations were checked against numerical outputs, with post hoc human interpretation kept separate from agent-facing memory.

\subsection{Cross-campaign abstraction and mechanical assessment}

Cross-campaign abstraction used three LLM synthesis steps. Campaign records were first organized into mechanism summaries connecting geometry, load redistribution, crack evolution, score components, and counterevidence. These summaries were synthesized into conditional principles with predicted signatures and falsification tests, then compressed into a textbook with at most four core entries while retaining evidence links. The prompts supplied standard mechanics vocabulary and requested observable signatures, failure conditions, and tests for later FEM campaigns. The saved textbook contains four entries linked to nine source principle topics, with design practices and diagnostic warnings. It was retained as external guidance for Stage II.

Figure 5A summarizes all 270 evaluated Stage I architectures. Reinforcement occupancy gives the fraction containing reinforcement in each design cell. Crack activity gives the fraction in which each finite element was deleted by the end of loading, evaluated on the common undeformed mesh without opacity weighting. One reaction history was plotted for each architecture, preserving all accepted states without smoothing or resampling. Reaction is the sum of the magnitudes of the upper and lower boundary forces; field position denotes the center of the traveling opening profile.

Representative Stage I cases in Fig.~\ref{fig:5}B--D were assessed using reinforcement layouts, final fracture maps, and accepted-state reaction and fracture records. For Fig.~\ref{fig:5}B, element-failure events during a plateau in precrack-associated extension support an interpretation of main-crack arrest with daughter-crack formation; the element-resolved nucleation sequence was not reconstructed. Table S2 separates the LLM's verbatim entry titles from our mechanical interpretation. Proposed falsification tests were not counted as completed validations. Pale red marks deleted elements adjoining retained shared nodes. This numerical connectivity can transmit load and was not interpreted as physical crack bridging.

\subsection{Generalization campaigns and response analysis}

Stage II used the shared design and evaluation procedure, with the abstraction textbook and retained Stage I experience supplied as guidance. The LLM was instructed to combine principles across motifs, assign their intended mechanical roles, and predict responses. The compiled candidate pool increased from 20 to 24 architectures per cycle, while the evaluation budget remained at 15 FEM simulations. Four campaigns comprised 32 cycles and 480 evaluations; evaluations preceding a campaign continuation were counted once.

Stage comparisons included all 270 Stage I and 480 Stage II evaluations; the larger Stage II total reflects longer campaigns at the same 15-evaluation cycle budget. Score distributions used common bins and were normalized separately within each stage, retaining repeated evaluations as separate events.

The control campaign used the Stage II workflow and retained its own feedback and memory. Prior FEM examples, Stage I evidence and the learned textbook, and motif and benchmark guidance were omitted; generic instructions covered the task, scoring, validity checks, and diagnostic comparisons. Comparison used the first three cycles of the control and four Stage II campaigns (45 FEM evaluations per campaign).

\subsection{Supplementary movie visualization}

Movies S1 and S2 were rendered in OVITO 3.15.4 \citep{stukowski2010ovito} from deformed FEM nodes attached to surviving elements, without displacement magnification. Configurations and reaction histories were synchronized by the traveling opening-field position without interpolating accepted simulation states. Marker size does not represent ligament width.

\section*{Acknowledgments}

\textbf{Use of generative AI:} During manuscript preparation, an LLM was used to assist with language editing and code revision. The authors reviewed and edited the output, verified the scientific content, and are responsible for the final manuscript.

\textbf{Author contributions:} BN designed the study, conducted the experiment, analyzed the results, drafted and wrote the paper.

\textbf{Competing interests:} The author declares no competing interests.

\textbf{Data and materials availability:} Code, curated campaign records, reported finite-element evidence, and executable replay notebooks supporting this study are available in the FracLearn repository at \href{https://github.com/GenerativeMechanicsLab/FracLearn}{\textnormal{https://github.com/GenerativeMechanicsLab/FracLearn}}.

\FloatBarrier
\bibliographystyle{unsrtnat}
\bibliography{references}
\clearpage
\setcounter{figure}{0}
\renewcommand{\thefigure}{S\arabic{figure}}
\section*{Supplementary Materials}
\begin{figure}[htbp]
\centering
\includegraphics[width=1.0\linewidth,height=0.68\textheight,keepaspectratio]{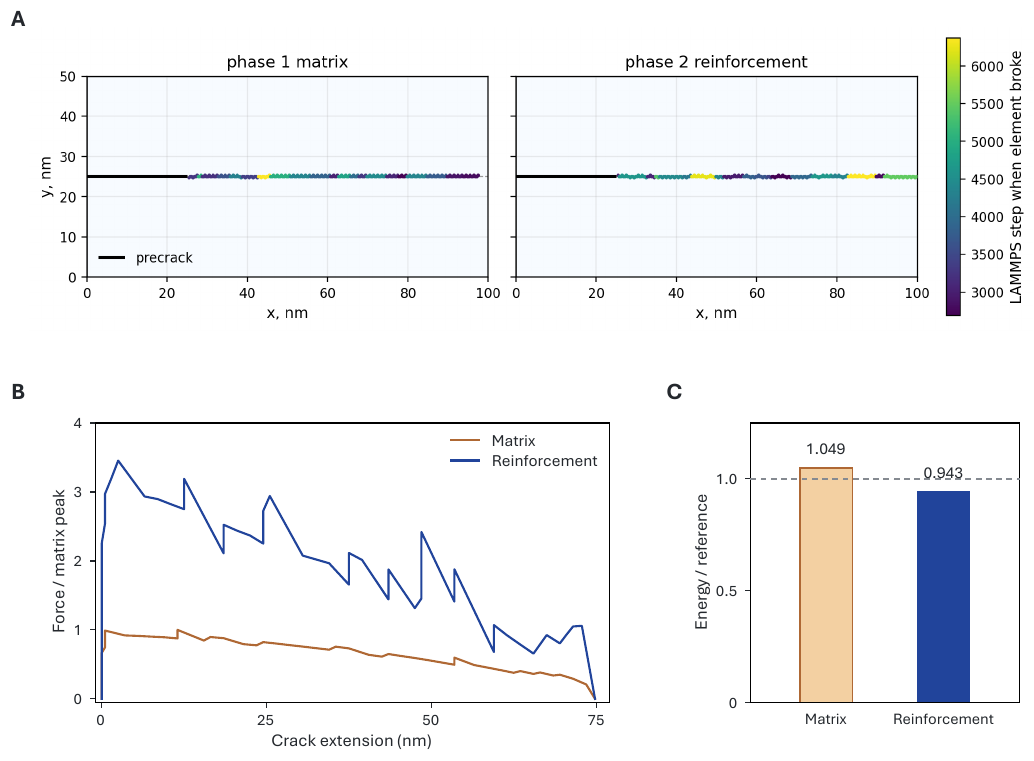}
\caption{\textbf{Homogeneous benchmarks of the fracture test.} (A) Final crack maps for matrix-only and reinforcement-only specimens under the prescribed surfing boundary condition. Black segments mark the precracks; colored points locate deleted-element centroids, with color indicating the recorded solver step at deletion. (B) Reaction force versus connected crack extension, normalized by the peak reaction force of the matrix specimen. (C) Energy consumed per projected crack area divided by the corresponding deletion-energy estimate; the dashed line denotes unity. The ratios are 1.049 and 0.943 for matrix and reinforcement, respectively. These benchmarks assess the selected mesh and loading protocol. Definitions and parameters are given in Materials and Methods.}
\label{fig:S1}
\end{figure}

\textbf{Table S1. Initial design hypotheses for the six experience campaigns.}

The entries condense the LLM's initial proposals and describe intended interactions and anticipated responses, rather than confirmed mechanisms. Channels, gates, and the local notch denote matrix-filled regions excluded from reinforcement; they are not additional precracks or voids. Campaign numbers match Fig.~\ref{fig:4}.

{\small\setlength{\tabcolsep}{4pt}\renewcommand{\arraystretch}{1.16}
\begin{longtable}{|
  >{\raggedright\arraybackslash}p{(\columnwidth - 6\tabcolsep - 4\arrayrulewidth) * \real{0.1758}}|
  >{\raggedright\arraybackslash}p{(\columnwidth - 6\tabcolsep - 4\arrayrulewidth) * \real{0.3890}}|
  >{\raggedright\arraybackslash}p{(\columnwidth - 6\tabcolsep - 4\arrayrulewidth) * \real{0.4352}}|}
\hline
\begin{minipage}[b]{\linewidth}\raggedright
\textbf{Campaign and motif}
\end{minipage} & \begin{minipage}[b]{\linewidth}\raggedright
\textbf{Architectural hypothesis}
\end{minipage} & \begin{minipage}[b]{\linewidth}\raggedright
\textbf{Anticipated fracture response}
\end{minipage} \\ \hline
\endhead
\endlastfoot
\textbf{1 --- Midspan stagger} & A broad interior reinforcement region near midspan, with variations in horizontal placement and staggering. & Central interception may promote repeated deflection and segmented crack advance while limiting reliance on boundary-adjacent reinforcement. \\ \hline
\textbf{2 --- Protected channel} & Two interior reinforcement lobes separated by a reinforcement-free matrix channel, with varied alignment and channel width. & Interaction with the lobes may increase crack tortuosity and distributed crack formation while preserving a matrix route through the central region. \\ \hline
\textbf{3 --- Terminal hook} & An asymmetric reinforcement region in the downstream half, organized as a diagonal chain. & Delayed engagement may produce a late kink or hooked crack path, increasing path length without extensive off-path damage. \\ \hline
\textbf{4 --- Left-entry ladder} & An oblique reinforcement chain extending from the crack-entry region toward midspan. & Earlier engagement may guide the crack through successive local kinks while maintaining connected advance through the reinforced region. \\ \hline
\textbf{5 --- Split-band gate} & Upper and lower interior reinforcement bands arranged around a narrow matrix gate. & The crack may pass the gate and engage the offset bands sequentially, promoting reorientation and distributed crack formation. \\ \hline
\textbf{6 --- Diagonal + notch} & An interior diagonal reinforcement chain interrupted near its midpoint by a small matrix-filled notch. & The interruption may promote staged engagement of upstream and downstream features, with directional changes near the notch instead of one uninterrupted barrier. \\ \hline
\end{longtable}
}

\textbf{Table S2. LLM-generated textbook principles and their mechanical interpretation.}

The first column preserves the LLM-generated entry titles verbatim. The remaining columns connect them to representative Stage I evidence and fracture mechanics. The interpretations apply to the present material model, loading, and objective; matrix occupies all unreinforced regions.

{\small\setlength{\tabcolsep}{4pt}\renewcommand{\arraystretch}{1.16}
\begin{longtable}{|
  >{\raggedright\arraybackslash}p{(\columnwidth - 8\tabcolsep - 5\arrayrulewidth) * \real{0.2266}}|
  >{\raggedright\arraybackslash}p{(\columnwidth - 8\tabcolsep - 5\arrayrulewidth) * \real{0.2238}}|
  >{\raggedright\arraybackslash}p{(\columnwidth - 8\tabcolsep - 5\arrayrulewidth) * \real{0.2720}}|
  >{\raggedright\arraybackslash}p{(\columnwidth - 8\tabcolsep - 5\arrayrulewidth) * \real{0.2776}}|}
\hline
\begin{minipage}[b]{\linewidth}\raggedright
\textbf{LLM wording (verbatim)}
\end{minipage} & \begin{minipage}[b]{\linewidth}\raggedright
\textbf{Mechanical interpretation}
\end{minipage} & \begin{minipage}[b]{\linewidth}\raggedright
\textbf{Representative evidence}
\end{minipage} & \begin{minipage}[b]{\linewidth}\raggedright
\textbf{Implication for design and assessment}
\end{minipage} \\ \hline
\endhead
\endlastfoot
P1. ``Guided main-crack interaction dominates simple obstruction'' & Control how reinforcement impedes and redirects the advancing crack, including arrest and associated daughter-crack formation. & Fig.~\ref{fig:5}B: left, main-crack arrest and separated fracture traces; right, a more continuous fracture route. Peak reactions are 3.65 and 3.26 nN, respectively. Additional element failure accompanies a force drop while the left main crack remains arrested. & Supports the mechanical value of arrest and renucleation \citep{hsueh2018renucleation}. The comparison does not establish that guidance generally outperforms obstruction. \\ \hline
P2. ``Realized interaction field and placement are first-order mechanics variables'' & Evaluate actual reinforcement placement, spacing, and alignment relative to the evolving crack and remaining load paths. & Fig.~\ref{fig:5}C: relocating six cells switches the fracture route around the downstream cluster; peak reaction changes from 2.65 to 4.11 nN despite 46 shared reinforcement cells. & Place reinforcement to engage the evolving crack route. Local geometry and alignment provide more useful mechanical guidance than a cluster-count or motif label alone \citep{guo2020tstress,brodnik2020guiding}. \\ \hline
P3. ``Secondary fracture is evidence, not a verdict'' & Interpret additional cracks using their accompanying fracture and load histories before assigning a mechanical benefit. & Fig.~\ref{fig:5}D: secondary traces coexist with final reactions of 0.087 and 0.865 nN despite similar peak reactions. & Assess how the location of secondary cracks relates to remaining load transfer. Similar fracture features can coexist with different residual responses; their presence alone does not determine their mechanical role. \\ \hline
P4. ``Metrics are diagnostic readouts with a strict hierarchy'' & Check rule validity, interpret fracture behavior, and assess the full score together with its component measures. & Fig.~\ref{fig:5}C: local placement changes fracture routing and peak reaction. Fig.~\ref{fig:5}D: secondary traces and similar peak reactions accompany markedly different final load retention. & Use complementary observables to assess the intended mechanism. Fracture-surface creation and effective toughness are distinct \citep{abid2019fracture,hossain2014effective}, and a combined objective retains trade-offs among responses. \\ \hline
\end{longtable}
}

\subsection*{Supplementary Movies}

\textbf{Movie S1. Crack--reinforcement interaction shapes fracture evolution.} Synchronized fracture evolution and total reaction for Stage I, Campaign 4, cycles 2 (A) and 3 (B), corresponding to Fig.~\ref{fig:5}B. Blue and peach denote reinforcement and matrix. Dashed lines mark the traveling opening-field center and the corresponding position on the reaction histories.

\textbf{Movie S2. Design revision restores crack engagement and load-bearing capacity.} Synchronized fracture evolution and total reaction for Stage II, Campaign 3, cycles 6 (A), 7 (B), and 8 (C), corresponding to Fig.~\ref{fig:6}C. Blue and peach denote reinforcement and matrix. Dashed lines mark the traveling opening-field center and the corresponding position on the reaction histories.

\end{document}